\documentclass[journal]{IEEEtran}
\usepackage{amsmath,amsfonts}
\usepackage{algorithmic}
\usepackage{algorithm}
\usepackage{array}
\usepackage[caption=false,font=normalsize,labelfont=sf,textfont=sf]{subfig}
\usepackage{textcomp}
\usepackage{stfloats}
\usepackage{url}
\usepackage{verbatim}
\usepackage{graphicx}
\usepackage{cite}
\usepackage{amsmath}

\usepackage{multirow}
\usepackage{multicol}
\usepackage{tabu}
\usepackage{color,soul}  
\usepackage{makeidx}
\usepackage{pdflscape}
\usepackage{booktabs}
\usepackage{hyperref}
\usepackage[capitalize]{cleveref}
\usepackage{color}
\usepackage{amssymb}
\usepackage{epstopdf}
\usepackage{url}
\usepackage{tabularx}
\usepackage{epstopdf}
\usepackage{balance}
\usepackage{float}
\usepackage{siunitx}
\usepackage{arydshln}
\usepackage{pifont}

\usepackage{url}
\usepackage{color,soul}
\usepackage{colortbl}
\usepackage[usenames,dvipsnames,svgnames]{xcolor}

\usepackage{etoolbox}
\usepackage[T1]{fontenc}

\begin{document}
%
\title{Can listening to more neighbours help CAVs be faster and safer?}
%
%
%

\author{Mohit~Garg
        and~Mélanie~Bouroche
\thanks{Mohit Garg and Mélanie Bouroche are with the School of Computer Science and Statistics, Trinity College Dublin, Ireland
        {\tt\small mgarg@tcd.ie, Melanie.Bouroche@tcd.ie}}%
}

\maketitle

\begin{abstract}

Connected Autonomous Vehicles (CAVs) are widely expected to improve traffic safety and efficiency by exploiting information from surrounding vehicles via V2V communication. A CAV typically adapts its speed based on information from the vehicle it follows. CAVs can also use information from vehicles further ahead within their communication range, and this results in improved traffic safety and efficiency. In mixed traffic scenarios, however, this may not always be possible  due to the presence of human-driven vehicles that do not have communication capabilities. Furthermore, as wireless vehicular networks are unreliable, information from other vehicles can be delayed or lost, which brings more challenges for CAVs in utilizing information from multiple leading vehicles. A few studies have investigated the impact of CAVs where they use information from multiple leading vehicles on traffic safety and efficiency,  but only in very limited scenarios (i.e., with a very small number of vehicles). 

In contrast, this paper investigates the impact of CAV car-following control  based on multiple leading vehicles information on both mixed traffic safety and efficiency in realistic scenarios in terms of imperfect communication, vehicle modelling, and traffic scenario. Results show that exploiting information from multiple, rather than a single, leading vehicles in CAV controller design  further improves both traffic safety and efficiency especially at high penetration rates. In addition to proper tuning of CAV controller parameters (control gains and time headways), the scale of the improvement depends on both market penetration rate (MPR) and communication reliability. A packet error rate (PER) of 70\%  leads to an increase in traffic efficiency by 4.18\% (at 40\% MPR) and 12.19\% (at 70\% MPR), compared to the simple single leading vehicle information based controller.
 
\end{abstract}


%
\IEEEpeerreviewmaketitle

\section{Introduction}

%
%
%
%

\textit{Connected Autonomous Vehicles (CAVs)} have the potential to enable improved traffic safety and efficiency by sharing information about themselves (such as their position, speed and acceleration) and various parameters of their traffic environment in real-time with other vehicles and infrastructure, so that they have a better situation awareness of surrounding vehicles and  events such as accidents and traffic jams that are out of the range of their on-board sensors~\cite{c1}. To achieve improved traffic safety and efficiency, CAVs employ cooperative car-following control algorithms to control vehicle speed in the longitudinal direction~\cite{c2}. The information flow topology (IFT) specifies the vehicles from which a CAV uses information to make decisions~\cite{c3}. These include  predecessor-following (each vehicle receives information from the vehicle in front of it), multiple-predecessor-following (MPF - each vehicle receives information from multiple vehicles in front of it), bidirectional topology (each vehicle receives information from the vehicle in front and the vehicle in back of it), etc. In fully connected environments, CAVs can  adapt their speed based on information from one (predecessor-following IFT) or more (MPF IFT) leading vehicles, via V2V communication within their communication range~\cite{c4}.  It is well demonstrated that the minimum time headway (expressing the desired inter-vehicle distance) is dependent on the number of leading vehicles and increasing the number of leading
vehicles information in CAV car-following control design can provide better traffic efficiency without compromising safety~\cite{c5,c6,c4}.

While CAVs have the potential to improve traffic performance in a fully connected environment, they face additional challenges in their driving when they interact with Human Driven Vehicles (HDVs) in mixed-traffic scenarios (i.e., comprised of HDVs and CAVs) due to the uncertainty in human's driving behavior e.g., larger reaction times, perception errors, etc.~\cite{c7}. Another challenging problem in utilizing CAVs technology is the unreliability of wireless communication networks. Delays and packet losses in communication links due to dense traffic, communication interference, channel fading, etc. may jeopardize CAVs driving behavior and negatively affect traffic performance in terms of reduced traffic safety and efficiency~\cite{c8}. 

Many control approaches that address the potential challenges of communication impairments and uncertainties present in the driving behavior of human-driven vehicles in mixed traffic environment have been reported in the literature~\cite{c9,c10}. Most of them, however, have been designed based on the simple predecessor-following IFT only, and degrade CAVs' control mode to ACC if information is not available from the preceding vehicle due to communication failures or it being a HDV, thereby resulting in decreased traffic safety and efficiency. Although traffic safety can be achieved by increasing the time headway, this is at the expense of reduced traffic efficiency~\cite{c11}. Furthermore, in most studies, CAVs have been assumed to operate in either realistic traffic scenarios with perfect communication, or with only a very small number of vehicles with imperfect communication. In recent years, a few researchers have attempted to develop cooperative car-following control algorithms based on the MPF IFT, i.e., considering information from more than one leading vehicle, for CAVs operation in unreliable communication and/or mixed traffic environment~\cite{c12,c13}. These studies, however, have been performed in very limited scenarios only (e.g., assuming a platoon formation with a designated leader and follower vehicles). 

 Time headway, control gain parameters, max acceleration and deceleration, sensor and actuator delays, wireless network condition, penetration rate, traffic demand and road network type, can all affect traffic safety and efficiency. Due to the large number, and complexity, of these factors, existing studies generally focus on a subset of them only, e.g., only a single type of time headway is evaluated, the impact of CAVs on traffic safety and efficiency is evaluated only on one type of a controller parameters, safety and efficiency are evaluated separately, or simplifying assumptions about the traffic scenario are made. Our study extends existing studies by investigating the impact of CAV control based on multiple leading vehicles information on both traffic safety and efficiency, in realistic communication and traffic flow scenarios at different penetration rates, using real traffic data of an Irish motorway (the M50 motorway, in Ireland), and then evaluates the effect of controller parameters (i.e., control gains and time headways) on traffic safety (time-to-collision (TTC)) and traffic efficiency (travel time) performance. 
 
 The contributions of this paper include: (1) designing a longitudinal controller for CAVs that can utilize information of multiple vehicles ahead in mixed traffic scenarios, (2) tuning of controller parameters (control gains and time headways) to make it robust against communication failures and to maintain a good trade-off between traffic safety and efficiency, and (3) investigating the advantages of exploiting information from multiple, rather than a single leading vehicle  on traffic safety and efficiency, using the designed CAV controller under different penetration rates of CAVs in realistic scenarios in terms of imperfect communication and traffic flow.

The paper is organized as follows: Section~\ref{sec:related_work} reviews the related work exploring CAV control  using multiple leading vehicles information, and their impact on traffic safety and efficiency. Section~\ref{sec:MPF_control} presents the MPF topology-based controller design along with the tuning procedure of its parameters. Section~\ref{sec:evaluation} presents the simulation set up and evaluation scenarios based on the Plexe simulator. Section~\ref{sec:results} analyzes the simulation results of different scenarios and discusses them in detail. Finally, Section~\ref{sec:conclusion} concludes the paper and discusses the future scope of the work.  

 

\section{Related Work}\label{sec:related_work}

Most existing research investigating the impact of CAVs on mixed traffic safety and efficiency in realistic scenarios has assumed that CAVs use information from their immediate leading vehicle only. Conversely, the advantages of multiple leading vehicles information over a single leading vehicle information in CAV controller design have been widely studied in the literature, in fully connected environments. These studies show that exploiting information of multiple leading vehicles in CAV controller design allows them to drive at smaller time headways, thereby resulting in a higher traffic efficiency~\cite{c5,c4,c6}. 

In mixed traffic, however, some vehicles are human-driven vehicles that do not broadcast information. This limits the  ability of CAVs to communicate with multiple vehicles, hindering them to utilize their full potential~\cite{c14,c15}. In such scenarios, some existing studies propose that HDVs should be equipped with wireless communication devices so that CAVs can exploit information from all leading vehicles within their communication range~\cite{c10,c16,c17,c18}. But, the more information is transmitted, the higher the likelihood of communication delays, packet losses, etc., due to the congestion of communication channels~\cite{c19,c20}. In contrast, a few studies investigated CAVs' control algorithms designed based on complex IFTs other than the simple predecessor-following IFT, without equipping communication devices in HDVs~\cite{c12,c13}. \cite{c12} studied a CACC system exploiting information from multiple-leading vehicles in mixed traffic, and a strategy is designed to determine whether the data received  from other vehicles is incorporated into the car-following control algorithm. However, this study was limited to a platoon of 8 vehicles only, with a fixed sequence of HDVs and CAVs. Furthermore, \cite{c13} developed a consensus-based control model for mixed traffic platoons using the variable time-headway (VTH) policy and the leader-predecessor-following IFT, assuming that the platoon leader is a CAV. The stability conditions of the mixed vehicle platoon system are derived through analytical study, and simulation experiments are conducted under different penetration rates, vehicle ordering and reaction times. Results show that the car-following control model designed based on the VTH spacing policy is better in terms of string stability of a mixed vehicle platoon than the control based on the constant time headway (CTH) policy. Furthermore, the effect of different penetration rates, vehicle ordering and reaction times on mixed platoon stability are shown through simulation studies. Results indicate that increasing the penetration rate of CAVs in the mixed vehicles platoon has a positive impact on the platoon stability. Simultaneously, CAVs in front of HDVs can help HDVs to better track the leader’s state changes. Furthermore, increasing the human driver's reaction time has a negative impact on mixed platoon stability. Another study claims that, for a mixed traffic platoon with multiple-predecessor following IFT, the optimal platoon control performance is obtained when all HDVs move behind all CAVs~\cite{c3}.

The aforementioned studies, however, assume perfect communication. In practice, communication delays and packet drops might affect the results. Very few studies have analyzed the impact of CAVs on traffic performance in the presence of communication impairments i.e., delays and packet losses~\cite{c10,c17}. The study presented in~\cite{c10} developed a multiple-predecessor following IFT-based cooperative control strategy for CAVs to enhance mixed traffic flow dynamics in the presence of time-varying delays in the V2V communication links and human-driver reactions. They showed that by properly tuning control gains, CAVs can achieve improved traffic efficiency by exploiting information from multiple leading vehicles even in the presence of communication imperfections. Furthermore, Zhang~\textit{et al.} proposed the connected cruise control strategy, which exploits information from multiple leading vehicles ahead via V2V communication in order to achieve string stability in the presence of fixed delays~\cite{c17}.

\begin{table*}[htbp]
\caption{Control strategies based on different IFTs for CAVs operation in mixed traffic and their impact on traffic performance}
\begin{center}
\scalebox{0.8}
{
\footnotesize
\begin{tabular}{>{\centering}m{.13\textwidth}>{\centering}m{.08\textwidth}>{\centering}m{.06\textwidth}>{\centering}m{.12\textwidth}>{\centering}m{.1\textwidth}>{\centering}m{.09\textwidth}>{\centering}m{.09\textwidth}>{\centering}m{.11\textwidth}>{\centering}m{.07\textwidth}m{.07\textwidth}>{\centering\arraybackslash}m{.08\textwidth}}
\noalign{
\hrule height 1.3pt
}%

\centering
\textbf{Reference}&\textbf{Vehicle types}&\textbf{MPR (\%)}&\textbf{Time headway settings}&\textbf{Control strategy}&\textbf{Road network}&\textbf{Traffic scenario}&\textbf{Realistic communication}&\textbf{Traffic efficiency}&\textbf{Traffic safety}\\
\hline
Zhang~\textit{et al.}~\cite{c12}& CAVs, HDVs, Connected HDVs &  -& 1 and 1.2s & CAV- Selective MPF IFT-based CACC, HDV- OVM, Connected HDV- IDM &A string of 8 vehicles & - & \checkmark& \ding{56}&\checkmark\\
\hline
Zhang~\textit{et al.}~\cite{c17}& CCC-equipped, Connected HDVs &  -& 3 and 4m (CAVs), 38 and 40m (HDVs) & Multiple vehicles ahead information-based connected cruise control & Single-lane& A platoon of four vehicles& \checkmark&\ding{56}&\ding{56}\\
\hline
Vaio~\textit{et al.}~\cite{c10}& CAVs, Connected HDVs& - & Distance headway-20m & Multiple vehicles ahead information-based CACC & Single-lane & String of three, eight and twenty vehicles & \checkmark& \checkmark&\ding{56}\\
\hline
Rahman~\textit{et al.}~\cite{c14}&CAVs, AVs, CVs, HDVs & 0, 20, 40, 60, 80, 100& 1.1s (HDVs), 0.6s (AVs and CAVs) & Multiple-predecessor-following IFT-based CACC& 14-miles long freeway & Congested & \ding{56}&\checkmark&\checkmark\\
\hline
Cui~\textit{et al.}~\cite{c18}& CAVs, HDVs, Connected HDVs &  10, 20, 80, 90& 1s & CAV- Multiple-predecessor-following and multi-time step-based CACC, HDV-OVM &A platoon of 12 to 16 vehicles & Congested & \ding{56}&\ding{56}&\ding{56}\\
\hline
Chen~\textit{et al.}~\cite{c13}& CAVs, HDVs &  0, 2, 4, or 6 HDVs & 1.2s (HDVs), 1s (CAVs)  & Leader-predecessor-following IFT-based adaptive consensus control &A platoon of 8 vehicles & Free-flow & \ding{56}& \ding{56}&\ding{56}\\
\hline
Wang~\textit{et al.}~\cite{c16}& CAVs, Connected HDVs &  - & 5m & Multiple-predecessor and successor IFT-based LCC &A platoon of 6 vehicles & - & \ding{56}& \ding{56}&\ding{56}\\
\hline
Avedisov~\textit{et al.}~\cite{c6}& CAVs, HDVs, Connected HDVs & 25, 50, 75, 100 & 45-55m (for HDVs), 29-45m (CAVs) & CAV- Connected cruise control, HDV- OVM with time delay  & Single-lane & String of hundred vehicles &\ding{56}&\checkmark & \ding{56}\\
\hline
Ding~\textit{et al.}~\cite{c15} &CAVs, HDVs &0, 20, 40, 60, 80, 100& 1.5s (for HDVs), 0.6,1,1.5,2s (for CAVs) & CAV- IDM and SSA-ENN combined with IDM, HDV- IDM & A String of 20 and 40 vehicles & Free-flow, moderate and congested &\ding{56} &\checkmark& \checkmark\\

\noalign{
\hrule height 1.3pt
}%
\end{tabular}}
\begin{flushleft}
\checkmark: considered, \ding{56}: not considered.
\end{flushleft}
\label{tab:mixed traffic_2}
\end{center}
\end{table*}

Table~\ref{tab:mixed traffic_2} summarizes the different control strategies based on information from multiple vehicles ahead, and the impact of CAVs on traffic safety and efficiency, at different penetration rates and traffic scenarios. It shows that different IFTs such as predecessor-following (PF), two-predecessor-following (TPF), bidirectional (BDL), multiple-predecessor following (MPF) have been implemented for CAVs' controller design, but with a small number of vehicles only~\cite{c21,c19,c5}. Most existing studies based on MPF IFT-based CAV control in mixed traffic have assumed a fully connected environment by equipping HDVs with wireless communication devices (i.e., connected HDVs), so that CAVs can exploit information from all leading vehicles within their communication range. However, it is unlikely that all HDV will be retrofitted with communication devices in a short period of time, consequently, the mixed traffic environment comprising of only HDVs and CAVs will exist~\cite{c15}. In addition, most studies are limited to perfect communication. Furthermore, they focus on analyzing the impact of CAVs on string stability, rather than evaluating the impact on traffic safety and efficiency. To fill this research gap, this paper investigates the impact of car-following control algorithms of CAVs with multiple leading vehicles information on traffic safety and efficiency in realistic scenarios in terms of vehicle modeling, traffic scenario and communication. Furthermore, this paper investigates the effects of the controller gain parameters and time headways on both traffic safety and efficiency.

\section{Multiple-Predecessor-Following (MPF) IFT-based Control}\label{sec:MPF_control}

In this paper, we aim to design a multiple-predecessor-following IFT-based car-following control algorithm for CAVs where they can adapt their speed based on information from multiple-leading vehicles within their communication range. To regulate the inter-vehicle distance and speed error to a minimum value, this controller takes the distance and speed errors of multiple leading vehicles as the feedback signals~\cite{c27}. The \textit{constant time headway (CTH)} policy is employed in the car-following control design, as it is most suitable for general information flow topologies such as multiple-predecessor-following topology~\cite{c5}. In this spacing policy, the desired time-headway is fixed, however, the desired inter-vehicle distance varies with the vehicle velocity~\cite{c23}. 

\subsection{Controller design}\label{ssec:controller}

We design a CAV longitudinal controller that incorporates multiple leading vehicles information in its algorithm, for mixed traffic scenarios with no fixed information flow topology. This controller allows the incorporation of multiple leading vehicles information (only if they are CAVs) without assigning a designated leader and follower vehicles, thereby making it more flexible and scalable for its implementation in realistic traffic scenarios.  In addition, we design our controller such that it still functions in mixed traffic, including when all neighbour vehicles are HDVs that do not have communication capabilities. In that case, a CAV degrades its car-following control mode to a sensor-based control.

The longitudinal control dynamics of vehicle \(i\) are represented as~\cite{c3},

\begin{equation}\label{eq:1}
\begin{split}
\dot{x_i}(t)&={v_i(t)},\\
\dot{v_i}(t)&={u_i(t)}=f_i^c(v_i(t), \Delta x_{i,j}(t), \Delta v_{i,j}(t))
\end{split}
\end{equation}

 \noindent where \(x_i\) denotes the position of the rear bumper of vehicle \(i\), \(v_i\) and \(u_i\) are the  velocity and desired acceleration of vehicle \(i\), respectively. \(\Delta x_{i,j}\) and \(\Delta v_{i,j}\) are the position and velocity difference of vehicle \(i\) with respect to its neighbour vehicle \(j\).

 The desired distance between vehicle \(i\) and its neighbour vehicle \(j\) in the CTH policy under MPF topology is formulated as,

\begin{equation}\label{eq:2}
\begin{aligned}
{x_{ij}^d}=\sum_{j \in N_i^c }({h_{ij}v_{ij}+d_{ij}})
\end{aligned}
\end{equation}
 \noindent where \(h_{ij}\geq0\) and \(d_{ij}>0\) denote the time headway of vehicle \(i\) and desired standstill gap of vehicle \(i\) with respect to its neighbour vehicle \(j\), respectively.

The MPF-based linear control algorithm under CTH policy is denoted as~\cite{c3},

\begin{equation}\label{eq:3}
\begin{split}
{u_i^c (t)}=\sum_{j \in N_i^c }a_{ij}\{\alpha[x_j(t)-x_i(t)-x_{ij}^d(t)]+ 
\beta[v_j(t)-v_i(t)]\}
\end{split}
\end{equation}

 \noindent where controller parameters (\(\alpha\), \(\beta\)),  the weight factor of kinetic information (position and speed), are feedback control gains that needs to be properly tuned to achieve optimal traffic performance. \(N_i^c(t)\) and \(a_{ij}\) denote neighbour set of CAVs under MPF topology and varying information flow topology (\(a_{ij}\) is 1 when information is available from \(j\) to \(i\), otherwise 0), respectively.

\subsection{Controller parameters tuning}

The generality of the impact of CAVs can be challenged by a different set of parameters such as time headway, control gain parameters, sensor and actuator delays, wireless network condition, penetration rate, traffic demand and road network type, that can affect traffic safety and efficiency. For this reason, controller parameters design is  critical to the traffic performance. Therefore, this work investigates how the controller parameters affect the system performance and based on the analysis, how to select the optimal (control parameters) values to
enhance the traffic performance.

We performed the experiment design of tuning the controller parameters (control gains and time headways) of CAVs based on a validation network, at four different control gains configurations and different time headways, using the controller framework detailed in Section~\ref{ssec:controller}, then validated the tuned controller parameters on a realistic road network (the M50 motorway, in Ireland). In all control gains configurations, only  the velocity error coefficient~\(\beta\) is varied (between 1 and 4), while keeping the position error coefficient~\(\alpha>0\) as constant and equal to 1, similarly to related work~\cite{c5,c3,c2}. This is also due to the fact that in the CTH policy, the desired inter-vehicle distance varies according to the vehicle velocity only~\cite{c5}. Furthermore, in all configurations the time headway was varied between 0.4s and 1s in 0.2s increments. This range of headways between vehicles was chosen according to related work~\cite{c5,c14}.

\begin{figure*}[htbp]
	\small
	\centering
	\subfloat[PER 0\%]{\includegraphics[width=9cm, height=7cm]{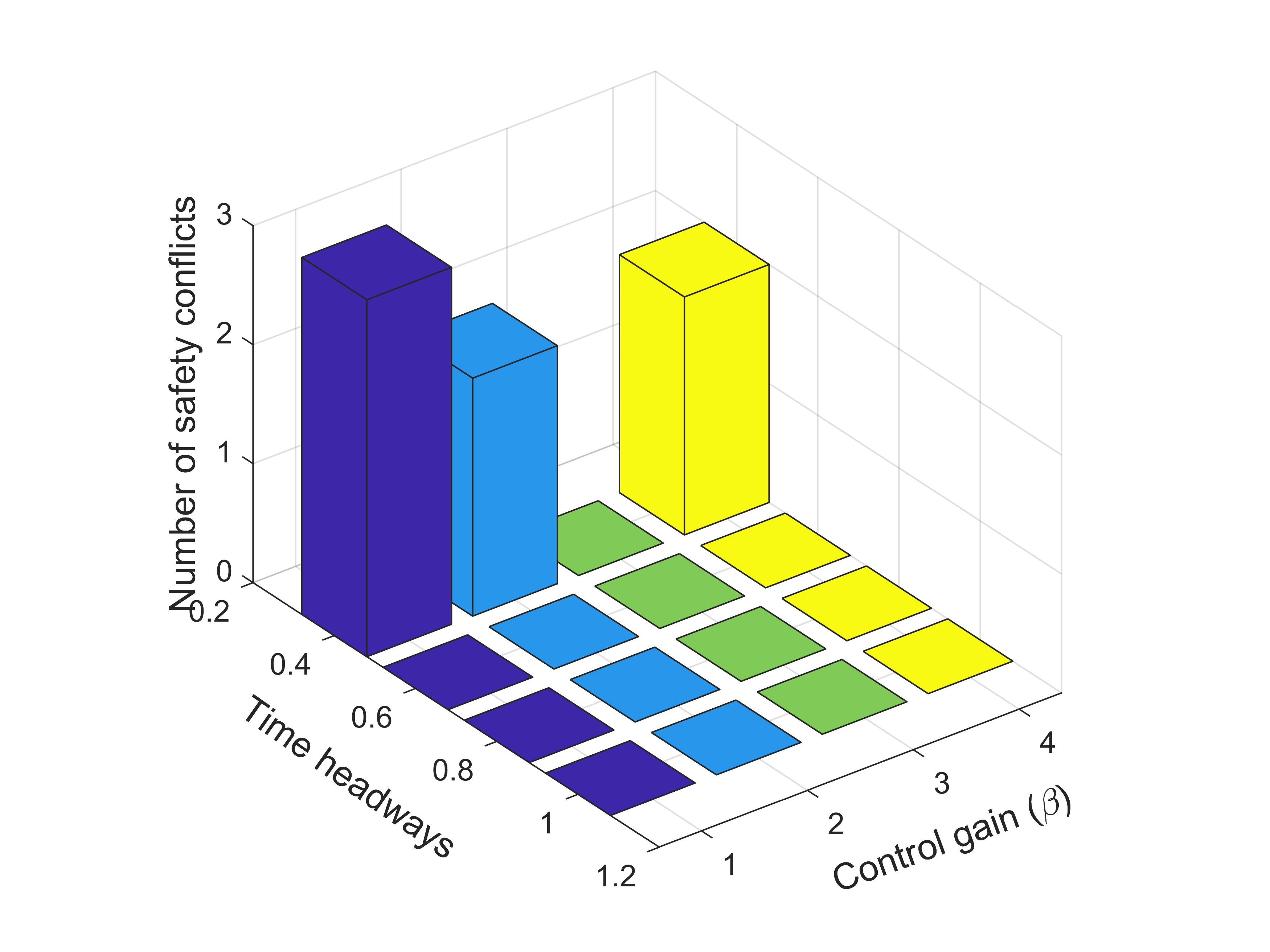}%
\label{fig:a1}}
\hfil
\subfloat[PER 70\%]{\includegraphics[width=9cm, height=7cm]{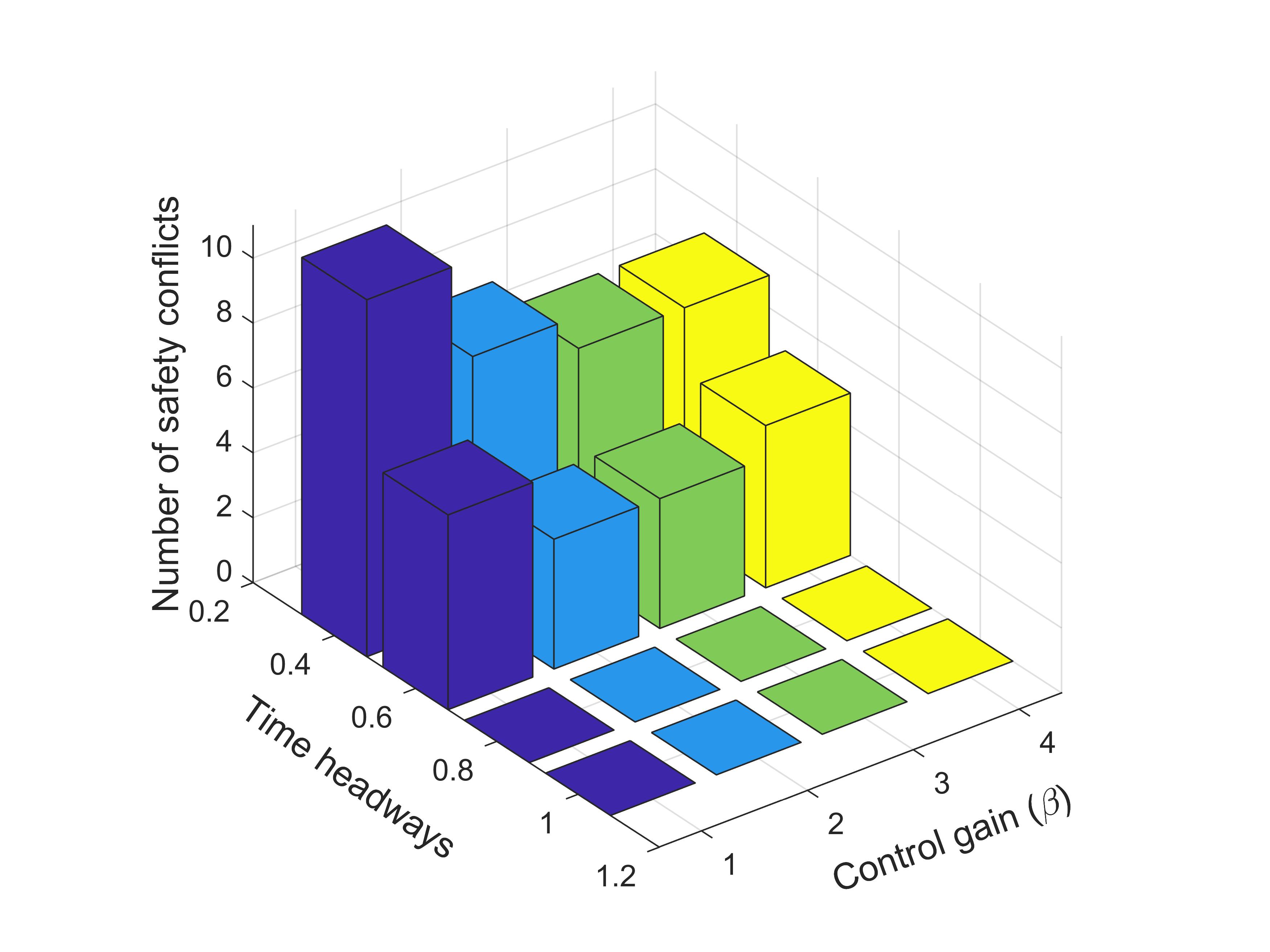}%
\label{fig:b1}}
\hfil
\subfloat[PER 0\%]{\includegraphics[width=9cm, height=7cm]{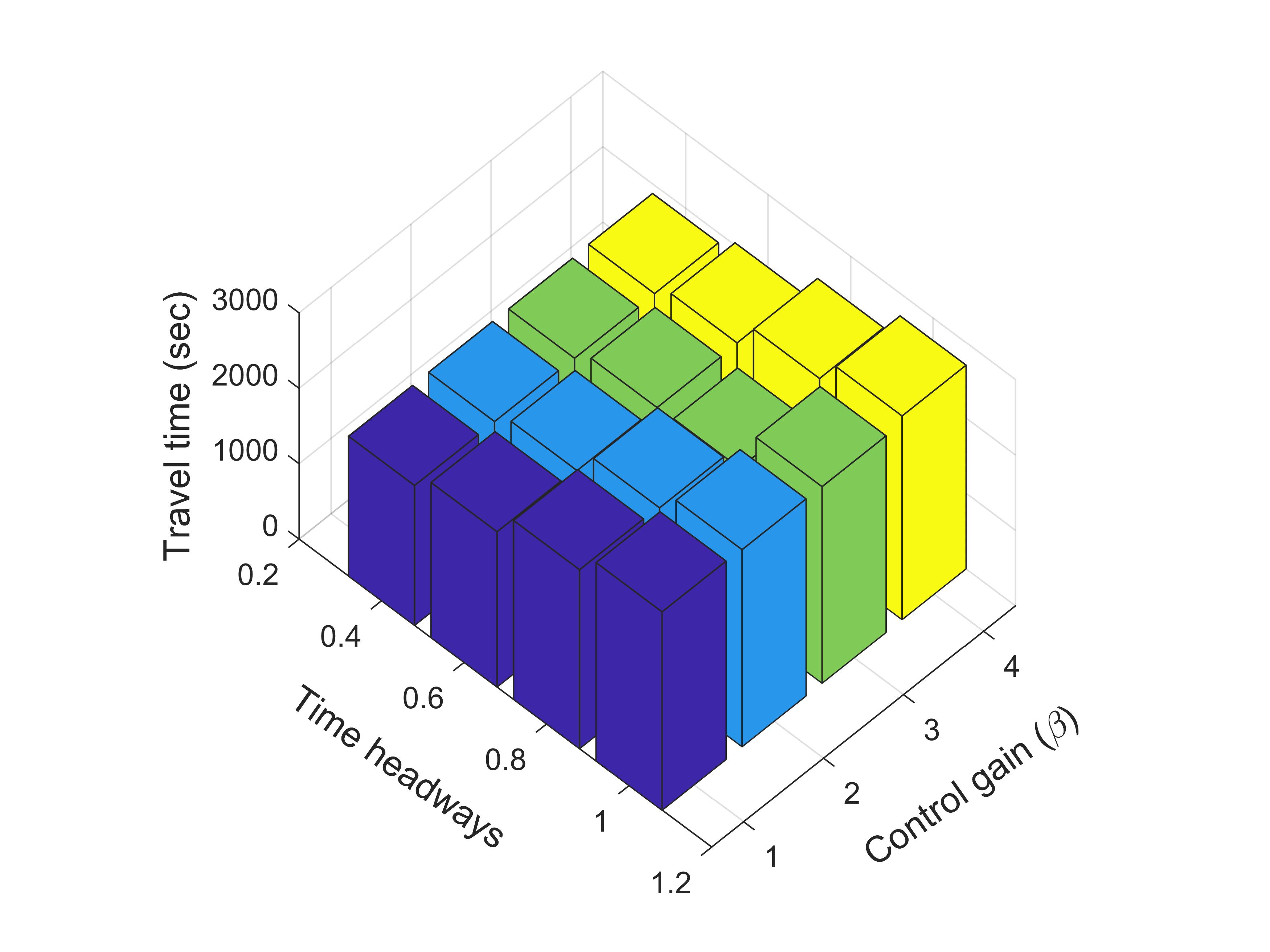}%
\label{fig:c1}}
\hfil
\subfloat[PER 70\%]{\includegraphics[width=9cm, height=7cm]{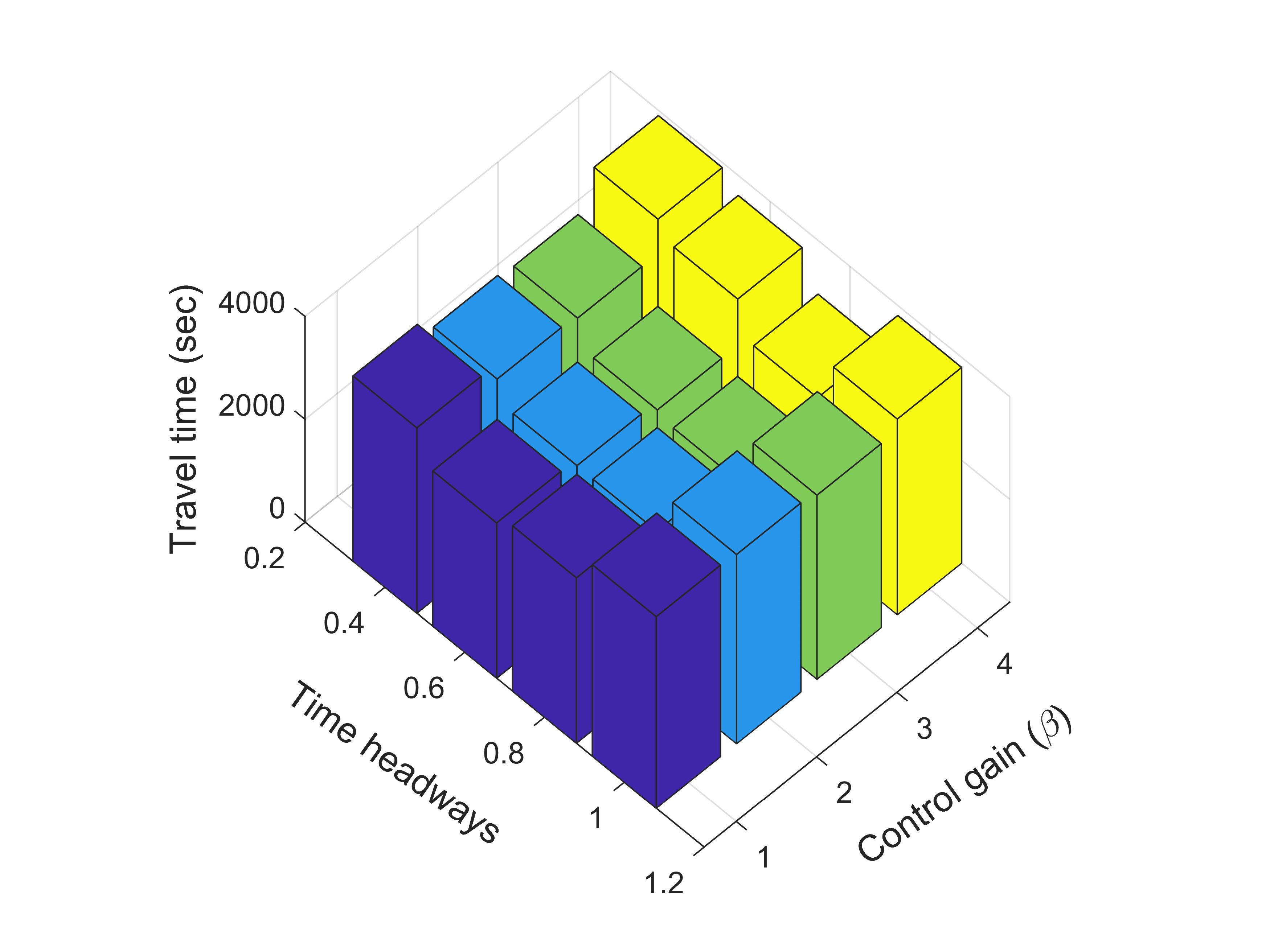}%
\label{fig:d1}}
\hfil
\caption{Number of safety conflicts and travel time at full penetration rate of CAVs for different control gains and different time headways, with and without packet drops}
\label{fig:ttc_tr}
\end{figure*}

Fig.~\ref{fig:ttc_tr} shows the 3D plot of traffic safety (i.e., the number of safety conflicts) and efficiency (i.e., travel time) at different headways and control gains, at 100\% penetration rate of CAV, with and without packet drops. It is observed that there is no significant effect of choosing different control gain values on traffic safety in perfect communication environments (Fig.~\ref{fig:ttc_tr}~(a)). The number of safety conflicts is zero for all time headways and control gains, except for very small time headway and high control gain values. Furthermore, it is observed that traffic efficiency can be improved significantly by adopting smaller time headways for all control gains (Fig~\ref{fig:ttc_tr}~(c)).  In imperfect communication environments, however, control gains have a significant effect on both traffic safety and efficiency. Traffic efficiency improves significantly with higher gain values and smaller time headways, but at the cost of reduced traffic safety (Fig~\ref{fig:ttc_tr}~(d)). Although traffic safety can be improved by increasing the time headways, this is at the expense of reduction in traffic efficiency. The above trade-off between traffic safety and efficiency can be resolved by selecting the optimal control gains for which the time headway value is smaller i.e., (control gains (\(\alpha\)=1, \(\beta\)=3), time headway (0.8s)). Using this combination of time headway and control gains, CAVs do not need to massively increase the time headway in order to compensate for the effects of communication impairments. These values are used in the following.

\section{Evaluation}\label{sec:evaluation}
The effect of multiple-predecessor-following IFT-based control of CAVs on both traffic safety and efficiency is evaluated by simulation in realistic communication and traffic scenarios with different market penetration rates of CAVs, using real traffic data of an Irish motorway.

This section first describes the simulation set up and evaluation scenarios used in this paper. It then discusses the vehicle and communication network modeling.
\subsection{Simulation Setup}

A common way in the literature to investigate the impact of CAVs is through traffic simulations (to simulate CAVs in realistic traffic scenarios),  integrated with real-time wireless network simulations (to establish information exchange within vehicles). This study uses the Plexe simulator, which is an extension of vehicular networking simulator (i.e., Veins) for cooperative car-following applications such as cooperative adaptive cruise control (CACC) and platooning~\cite{c24}. Veins is a popular open source simulation framework based on the SUMO traffic simulator and the OMNET++ network simulator. It offers simulating both realistic traffic scenarios along with realistic modeling of physical and medium access layers to emulate an unreliable wireless communication network~\cite{c25}. The Plexe simulator used in this work provides the state of the art CACC controllers and also has the capability to implement a user-defined control algorithm~\cite{c24}.

\subsection{MPF-IFT-based controller implementation}
CAVs can broadcast their information to other CAVs within their communication range, therefore each CAV can receive information from multiple preceding vehicles for their control algorithm design. While all existing CACC controllers implemented in Plexe assume the existence of a platoon (with a designated leader vehicle and follower vehicles), we design a multiple-predecessor IFT-based CACC control design for CAVs operation on a large-scale road network (where it is not possible to assign a leader vehicle and follower vehicles), by modifying the source code of the existing control law (i.e., consensus-based controller~\cite{c26}) on the SUMO side. With respect to V2V communication, on the OMNET++ side, CAVs can already receive packets from any other CAV, within their  communication range.

\subsection{Car-following models}
The IDM car-following model is used in this work to model the car-following behavior of a human-driven vehicle. IDM is the most widely used car-following model for road traffic simulations due to its realistic stop-and-go traffic behavior. It determines the acceleration/deceleration of an ego vehicle based on input information such as the speed of the ego vehicle, and the position and speed difference with respect to its preceding vehicle~\cite{c27}. 

The IDM model is represented as~\cite{c28}:
\begin{equation} \label{eq:4}
\dot{v}(t)=a [1- \left( \frac{v(t)}{v_f} \right)^4- \left( \frac{s(t)^*(v(t), \Delta v(t))}{s(t)} \right)^2]
\end{equation}
where \(v\) and \(\dot{v}\) represents the speed and acceleration of the ego vehicle, respectively; \(v_f\) is the free flow speed; \(a\) is the maximum acceleration; \(s^*\) and \(s\) are the estimated and actual distance between adjacent vehicles, respectively; and \(\Delta v\) is the speed difference between adjacent vehicles.

The estimated distance between adjacent vehicles is given as:
\begin{equation} \label{eq:5}
s(t)^*(v(t), \Delta v(t))=s_0 + \max \left[0, v(t)T+ \frac{v(t) \Delta v(t)}{2\sqrt{ab}} \right]
\end{equation}
where \(T\)  represents the time headway; \(s_0\) is the minimum gap at standstill; and \(b\) is the desired deceleration. The IDM model parameters are selected according to~\cite{c29}.  The time headway parameter value of 1.5~s is chosen, according to related work using the same car-following model~\cite{c30,c9,c29,c11}.

\subsection{Communication network models}

To obtain information from surrounding vehicles via V2V communication, each CAV is equipped with the IEEE 802.11p V2V communication protocol-based network interface card. A CAV sends the information (position and speed) to its neighbour vehicles, provided that it has a limited communication range, at 10 Hz frequency and uses the default network parameters used in~\cite{c24}. The information will be received without any delay or loss if those vehicles are within its communication range. An unreliable vehicular network can be modeled by inducing packet drops at the medium access layer~\cite{c25}. In this work, to simulate imperfect communication, the frameErrorRate parameter is used to create packet drops at the MAC layer. Two values of frameErrorRate (i.e., 0 and 0.7) are chosen to represent no packet drops and 70\% packet drops, respectively. The frame error rate of 0.7 represents a realistic value of packet drops in congested traffic scenarios, according to the work in~\cite{c31}.

\subsection{Traffic Scenario}

To evaluate the impact of CAV penetration on traffic safety and efficiency in realistic traffic scenarios, this paper simulates a section of the M50 motorway (a 7-km long, 4-lane road network), in Ireland with real traffic data originally created by Gueriau and Dusparic~\cite{c32}. This motorway is the most busiest road network in Ireland, with a very high volume of traffic (up to 25,316 vehicles) during the busiest time period (i.e., 07:00-08:00). A simulation time window of 30 minutes (07:00-07:30),  with a step-length of 0.1~s is chosen within that period. Considering the computational complexity of simulations, and that it takes over 5 days of real time to complete one simulation on a high performance computing cluster, a longer window was considered impractical. In the Veins simulator, a simulation parameter manager.firstStepAt is leveraged  to advance the simulation to a particular point in time when performing a simulation to analyze the impact in a particular time period. We choose the firstStepAt parameter value of 06:45 so that OMNET++ will start synchronising with SUMO at 06:45 allowing both road traffic and communication networks to be fully loaded in those 15 minutes before the simulation window~\cite{c11}.

\section{Results and Discussion}\label{sec:results}

Simulations are performed at five different market penetration rates of CAVs (i.e., 0\%, 20\%, 40\%, 70\%, 100\%) and two packet error rates (i.e., 0 and 0.7), resulting in a total of nine mixed traffic scenarios. Each scenario has been evaluated with two different control algorithms. In the first control algorithm, CAVs use information of  their immediate leading vehicle only (Predecessor-Following PF IFT) while in the second,  they use information from multiple leading vehicles (Multiple-Predecessor-Following MPF IFT).

The impact of CAV is evaluated using the following two metrics~\cite{c11}: 1) Travel time (to evaluate traffic efficiency), and 2) Time to collision (for the safety evaluation).

\subsection{Travel time}
The \textit{Travel time (TT)} gives information about the time vehicles take to travel a certain edge of the road network. We collected the mean speed for each edge of the road network using edge-based network state devices, and these  were divided by the length of each edge to get the travel time~\cite{c11}. The overall travel time is then obtained by adding up travel times for each edge in the simulated road network.

\begin{equation}
\text{TT\ (sec)}= \sum_i\frac{v_i}{q_i}
\end{equation}

 \noindent where \(v_i\) and \(q_i\) are the mean speed and length of the \(i\)th edge, respectively.
 
\subsection{Time to collision (TTC)}
\textit{TTC} is a widely used traffic safety indicator in car-following control applications. It refers to the remaining time before a rear-end collision between two adjacent vehicles if they travel in the same lane and maintain their speed.
 
 \begin{equation}\label{eq:6}
\text{TTC}(t)= \frac{x_{i-1}(t)-x_i(t)-l_{i-1}}{v_i(t)-v_{i-1}(t)}, v_i(t)>v_{i-1}(t)
\end{equation}
\(x_{i-1}\) and \(x_i\) are the position of the leading vehicle and the ego vehicle, respectively; \(v_{i-1}\) and \(v_i\) are the speed of the leading vehicle and the ego vehicle, respectively; and\(l_{i-1}\) is the length of the leading vehicle.

A low TTC value represents a risky traffic situation and hence a suitable TTC threshold value must be selected to distinguish between risky and safe traffic situations. Different threshold values are selected in existing studies varying between 1 to 3~s. Two different threshold values of 0.75 s and 1.5 s are chosen in this paper for CAVs and HDVs, respectively, as per related work~\cite{c32,c11}.

\subsection{Traffic efficiency results}

\begin{figure}[htbp]
\centering
\includegraphics[width=9cm, height=7cm]{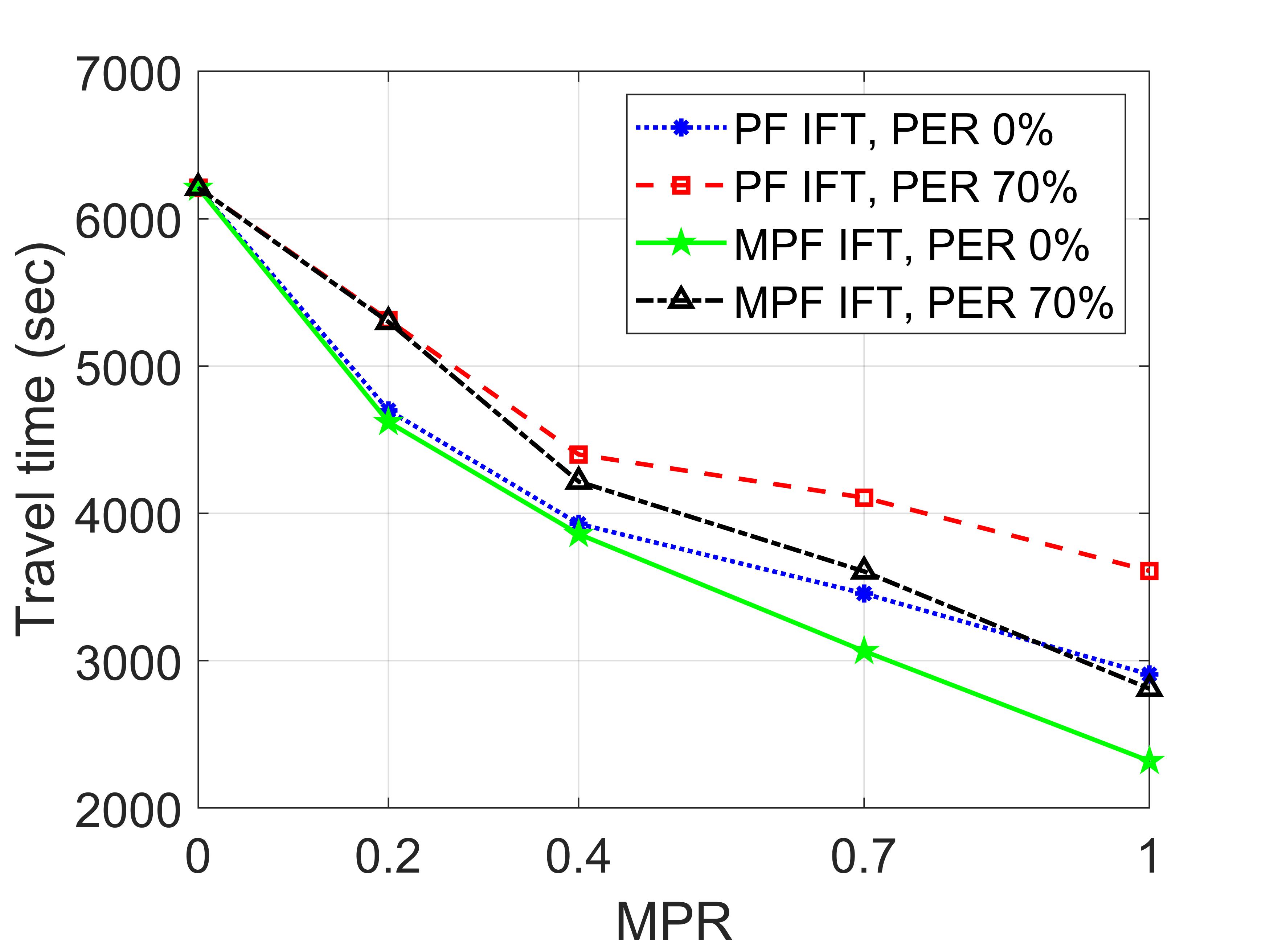}
\caption{Travel time at different penetration rates, packet error rates, \\and controllers}
\label{fig:TR_new}
\end{figure}

Fig.~\ref{fig:TR_new} shows the travel time at different penetration rates (0\%, 20\%, 40\%, 70\%, 100\%), with and without packet drops. It is observed that, as in other studies~\cite{c5,c4,c14}, in perfect communication environment (PER 0\%), traffic efficiency improves significantly when CAVs use information from multiple leading vehicles (MPF IFT) rather than a single leading vehicle (PF IFT) in their controller design, especially at high penetration rates. In the presence of packet drops (PER 70\%), however, a small decrease in traffic efficiency is observed at all penetration rates except the 100\% penetration rate, compared to results without packet drops. This is due to the fact that in imperfect communication environments, the improvement in traffic efficiency is not only dependent on increasing the number of leading vehicles, but also on proper tuning of time headway and control gains.

\subsection{Traffic safety results}

\begin{figure}[htbp]
\centering
\includegraphics[width=9cm, height=7cm]{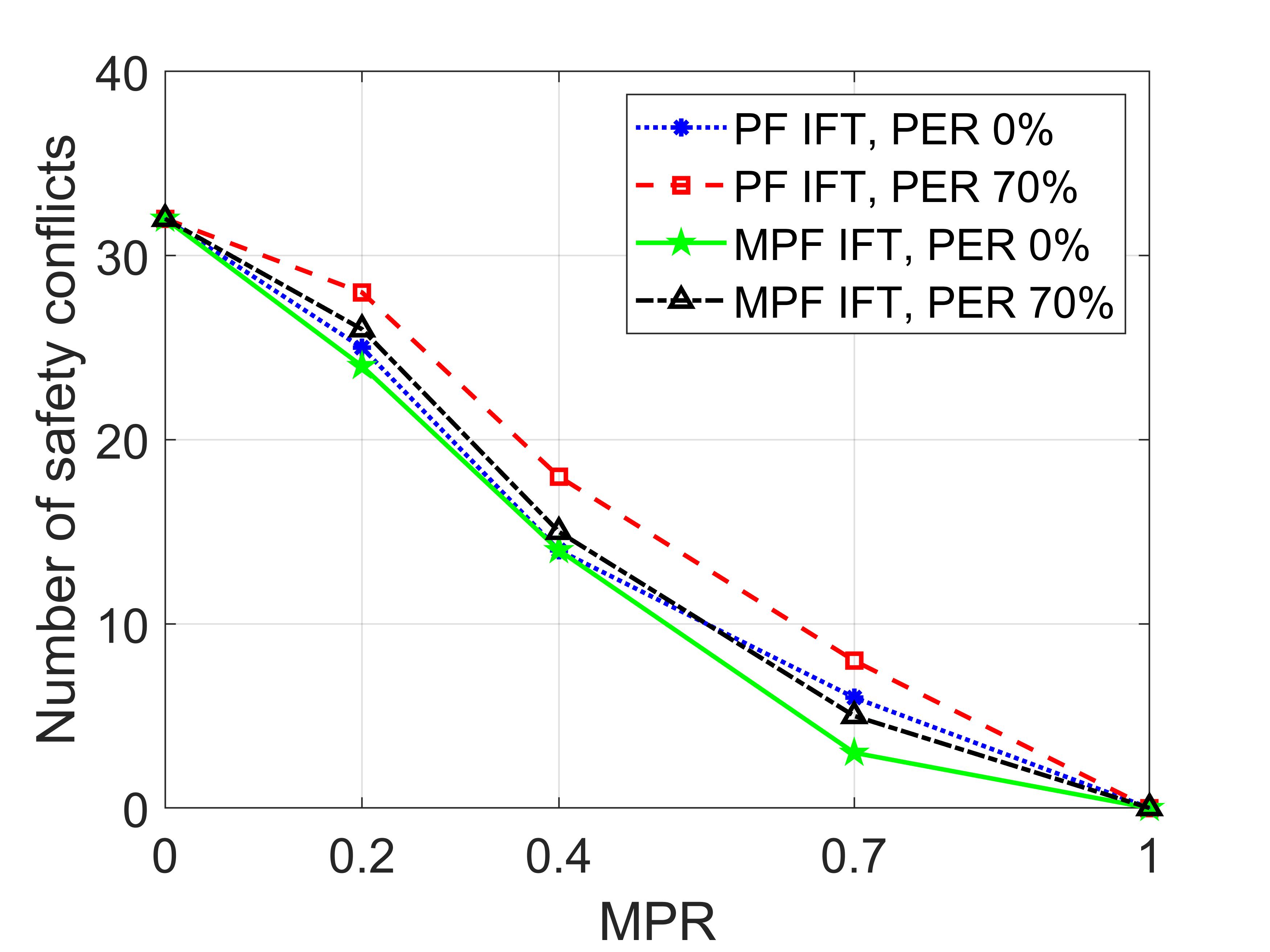}
\caption{Number of safety conflicts at different penetration rates, packet error rates, and controllers}
\label{fig:TTC_new}
\end{figure}

Fig.~\ref{fig:TTC_new} shows the number of safety conflicts based on time-to-collision (TTC) value at different penetration rates (0\%, 20\%, 40\%, 70\%, 100\%), with and without packet drops. It shows that when CAV controller uses information of multiple leading vehicles (MPF IFT) rather than the single leading vehicle (PF IFT), the number of safety conflicts decreases and this is more significant at high penetration rates. In the presence of packet drops, however, a substantial increase in the number of safety conflicts is observed at all penetration rates except the 100\% penetration rate for both single vehicle and multiple vehicles information based control, compared to results without packet drops. For both controllers, the number of safety conflicts is zero at 100\% penetration rate with packet drops due to proper tuning of control gains and time headways. Overall, CAVs provide a significant reduction in the number of safety conflicts when exploiting information of multiple leading vehicles in their controller design and the maximum reduction is found at high penetration rates. These results are consistent with the fact that by increasing the number of predecessors, CAVs can provide better traffic efficiency without compromising safety~\cite{c5,c4}.

\section{Conclusions and Future Scope}\label{sec:conclusion}

In this paper, we design a CAV controller based on multiple leading vehicles information in mixed traffic scenarios, and evaluate the impact of CAV penetration on both mixed traffic safety and efficiency in realistic scenarios in terms of imperfect communication and traffic flow scenarios on the busiest motorway in Ireland. Results show that in perfect communication conditions, CAVs can improve traffic safety and efficiency more effectively at high penetration rates,  when they use information from all preceding vehicles within their communication range rather than their immediate leading vehicle only. Furthermore, they show that imperfections in V2V communication links have adverse effects on both safety and efficiency. Traffic safety can be improved significantly by increasing the time headways, however, this is at the expense of further reduction in traffic efficiency. By properly tuning control gains, a small increase in CAV time headways results in improved traffic safety without compromising efficiency.

A  number of points remain to be addressed in future work. Firstly, the study needs to be extended to  road network types beyond highways (e.g., an urban network with signalized intersections) and to other vehicle types (e.g., heavy vehicles). Secondly, system uncertainties and hardware limitations (e.g., the actuator lags and sensor delays), should be modeled in the vehicle longitudinal dynamics. Thirdly, as aforementioned, controller parameters design is another issue critical to traffic performance. In this work, we have performed control gains tuning at 100\% CAV penetration rate only, but the effects of control gains on traffic safety and efficiency at different penetration rates need to be further investigated.

\end{document}